# *Ab initio* determination on the thermal evolution of the Earth's core

Wei-Jie Li[1], Zi Li[1], Yan-Bo Shi[1], Xian-Tu He[1,2], Cong Wang[1,2*], Ping Zhang[1,2*]

**Graphical Abstract**

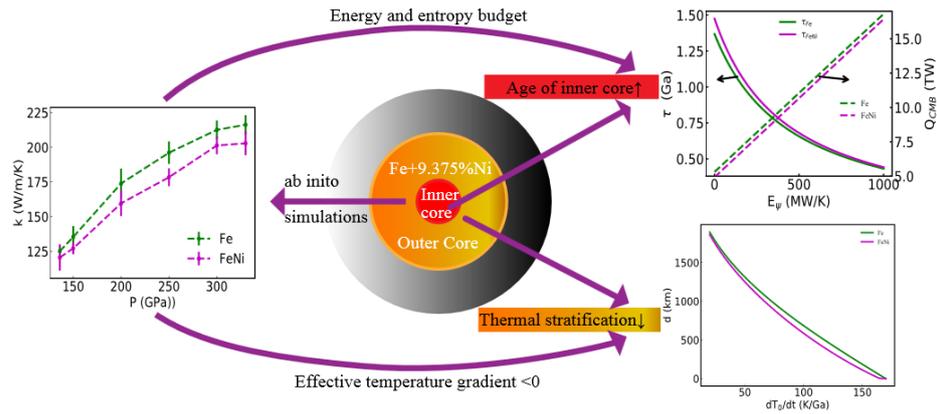

**Highlights**
- Thermal conductivity of FeNi mixtures at Earth's outer core via ab initio simulations.
- Age of inner core is 0.67 Ga when core-mantle heat flow is 12 TW.
- Ni reduces thermal stratification thickness to 320.12 km when $dT_o/dt$ = 126 K/Ga.

# *Ab initio* determination on the thermal evolution of the Earth's core

Wei-Jie Li[1], Zi Li[1], Yan-Bo Shi[1], Xian-Tu He[1,2], Cong Wang[1,2*], Ping Zhang[1,2*]

1 Institute of Applied Physics and Computational Mathematics, P.O. Box 8009, Beijing 100088, People's Republic of China

2 Center for Applied Physics and Technology, Peking University, Beijing 100871, People's Republic of China

Corresponding author: wang_cong@iapcm.ac.cn

Corresponding author: zhang_ping@iapcm.ac.cn

## Abstract

Earth's magnetic field is generated by the liquid outer core and sensitively depends on the thermal conductivity of the core. The dominant component of the Earth's core is Fe (~85%) and Ni (~10%). However, current estimates on FeNi mixtures have not been previously tested at high pressures. In this paper, ab initio simulations were first applied to calculations of the thermal and electrical conductivities of FeNi mixtures at Earth's outer core conditions. Compared with the results for pure Fe, the addition of Ni decreases the thermal conductivity (12.30 W/m/K on average) along the adiabatic curve in the outer core. Based on the restriction of the entropy production rate or Joule losses, the existence of Ni prolongs the age of the inner core. The age of the inner core is 0.66 Ga with pure Fe and 0.67 Ga with an FeNi mixture when heat flow at the core-mantle boundary is 12 TW. In contrast, we observe that Ni decreases the thickness of thermal stratification in the outer core by analyzing the effective temperature gradient. After inner core solidification, the thickness of thermal stratification is 417.02 km with pure Fe and 320.12 km with an FeNi mixture when the cooling rate at the core-mantle boundary is 126 K/Ga.

## 1. Introduction

The Earth's core is mainly composed of Fe, with the remainder consisting of Ni (~10%) (Vočadlo 2013) and other light elements (5 to 10%) (Nimmo 2015). Vigorous convection in the Earth's outer core powers the dynamo that sustains the magnetic field. Earth's outer core is an active area with the existence of thermal (Gubbins, Alfe et al. 2003) and compositional convection (Gubbins, Alfe et al. 2004). In a thermally advecting core, heat conducted along the core geotherm reduces the fraction of heat available to drive the geodynamo, which is sensitively dependent on the thermal conductivity of Earth's outer core. Previous studies have widely reported on related theories for the thermal and dynamic evolution of the Earth's core (Buffett, Huppert et al. 1996, Labrosse, Poirier et al. 2001, Gubbins, Alfe et al. 2003, Labrosse and Macouin 2003, Roberts, Jones et al. 2003, Gubbins, Alfe et al. 2004, Davies and Gubbins 2011, Deguen 2012, Davies, Pozzo et al. 2015, Nimmo 2015). The theory of the thermal evolution of Earth's core can infer the age of the inner core (Buffett, Huppert et al. 1996, Labrosse, Poirier et al. 2001, Gubbins, Alfe et al. 2003, Roberts, Jones et al. 2003, Gubbins, Alfe et al. 2004, Davies and Gubbins 2011, Nimmo 2015) and the thickness of the stable subadiabatic (or thermal) stratification in the outer core (Gubbins, Thomson et al. 1982, Lister and Buffett 1998, Gubbins, Alfè et al. 2015).

By applying an energy and entropy budget to the Earth's core, previous studies obtained the equation for the age of the inner core (Buffett, Huppert et al. 1996, Labrosse, Poirier et al. 2001, Gubbins, Alfe et al. 2003, Roberts, Jones et al. 2003, Gubbins, Alfe et al. 2004, Davies and Gubbins 2011, Nimmo 2015) based on several types of restrictions. Using restrictions on the core-mantle boundary heat flow (Labrosse, Poirier et al. 2001, Labrosse and Macouin 2003), the age is ~1.91 Ga for a 6.0 TW core-mantle boundary heat flow (Buffett, Huppert et al. 1996), whereas the age is ~1.00 ± 0.50 Ga at a low heat flow of 3.0 TW (Labrosse, Poirier et al. 2001). Based on restrictions from Joule losses (Roberts and Glatzmaier 2000, Roberts, Jones et al. 2003), the age is not more than 1.2 Ga in the absence of radioactive heating (Roberts, Jones et al. 2003). Thermal conductivity (Landeau, Aubert et al. 2017, Williams 2018) and heat flow across the core-mantle boundary (Roberts, Jones et al. 2003, Lay, Hernlund et al. 2008) provide implications for the structure, dynamics, and evolution of the core and deep mantle (Nimmo 2015). Previous studies have noted that the Earth's outer core is superadiabatic and had a low thermal conductivity during its early stages of formation (Buffett, Huppert et al. 1996, Labrosse, Poirier et al. 2001, Driscoll and Du 2019). However, high thermal conductivity results correspond to a younger inner core age, ranging from 0.30 to 1.80 Ga (Gubbins, Alfe et al. 2004), and a subadiabatic outer core, which has changed the view on the Earth's core (Pozzo, Davies et al. 2012, Gomi, Ohta et al. 2013, Ohta, Kuwayama et al. 2016). Based on restrictions for the entropy production rate with a high thermal conductivity, a heat flow of 7.0-17.0 TW implies an inner core age of ~0.60-1.30 Ga (Davies and Gubbins 2011, Pozzo, Davies et al. 2012, Nimmo 2015).

In contrast, the high thermal conductivity may also cause a subadiabatic condition in the outer core, with the appearance of stratification. Most seismic observation estimates of the thickness of the layers lie between 200 (Lay and Young 1990, Helffrich and Kaneshima 2013) and 300 km (Helffrich and Kaneshima 2010), but some are as large as 800 km (Kaneshima 2018). Modern observations of the geomagnetic field reveal fluctuations, such that predictions require 140-km-thick stratified layer with a heat flow of ~13 TW (Buffett 2014). Various theoretical mechanisms have been suggested to calculate the thickness of thermal stratification (Gubbins, Thomson et al. 1982, Lister and Buffett 1998, Helffrich and Kaneshima 2013, Hirose, Labrosse et al. 2013). The layer thickness is on the order of 100 km for a Nusselt number of Nu = 0.9 (Lister and Buffett 1998). Considering the temperature gradients (Davies and Gubbins 2011), the thickness ranges from 0 to 1,472 km, which depends on the specified parameters (Pozzo, Davies et al. 2012). An ~1,000 km stratified layer with the existence of an inner core can be inferred by the heat flux density (Gomi, Ohta et al. 2013). Adopting an approximate effective heat flux without the inner core, a stably stratified region occurs in the bottom two-thirds of the core (Nimmo 2015).

Recently, thermal conductivity has been an active area of research in the Earth Sciences. However, subsequent attempts to determine the thermal conductivity of Fe at Earth's core condition have produced discordant results (Konôpková, McWilliams et al. 2016, Ohta, Kuwayama et al. 2016). The thermal conductivity is ~30 W/m/K based on direct thermal conduction measurements by a laser-heated diamond-anvil cell (Konôpková, McWilliams et al. 2016) while the thermal conductivity of pure Fe is 90 W/m/K, as measured by the electrical resistance of Fe wires (Ohta, Kuwayama et al. 2016). A Shock experiment published in 2020 reported that the thermal conductivity around the core-mantle boundary condition is 60 to 130 W/m/K (Zhang, Hou et al.

2020). The thermal conductivity of liquid Fe$_{67.5}$Ni$_{10}$Si$_{22.5}$ has been inferred to be $88^{+29}_{-13}$ W/m/K at 140 GPa and 3,750 K (Ohta, Kuwayama et al. 2016). The electrical resistivity of hcp Fe-Ni alloys were measured using a four-terminal method in a diamond-anvil cell at up to 70 GPa at 300 K, which showed that resistivity increases linearly with an increasing Ni impurity concentration (Gomi and Hirose 2015). Based on high-pressure experiments of hcp Fe-Ni and Fe-Si alloys and related ab initio calculations, the thermal conductivity at the top of the Earth's core is 87.1 W/m/K (Gomi, Hirose et al. 2016). Previous studies have conducted electrical resistivity experiments on Fe-Ni systems at 4.5 and 8 GPa, showing that Fe-Ni alloys are more resistive than Fe by a factor of ~3 (Pommier 2020).

Since 2012, ab initio results have agreed with the high thermal results (144-223 W/m/K) (de Koker, Steinle-Neumann et al. 2012, Pozzo, Davies et al. 2012). The entire shocking Hugoniot curve of pure Fe was also reported (Wang, Zhang et al. 2014). When considering electron-electron scattering (Xu, Zhang et al. 2018, Zhang, Hou et al. 2020), the thermal conductivity of pure Fe is 147 W/m/K at the inner-core boundary and 97 W/m/K at the core-mantle boundary. The addition of light elements mainly decreases thermal conductivity (de Koker, Steinle-Neumann et al. 2012, Gomi, Ohta et al. 2013). However, although Ni is the second main element in the Earth's core, related theoretical calculations on the effects of Ni are limited. The transport properties of the Fe-Ni mixture throughout the entirety of Earth's core are still unknown. There is therefore a need for the direct determination of the thermal conductivity of the FeNi liquid mixture at pressures and temperatures characteristic of Earth's outer core.

In this paper, we theoretically investigate the electrical transport property of FeNi mixture at Earth's core condition. In the first case, the thermal and electrical conductivity of the FeNi mixture throughout the entirety of Earth's core were calculated by simulations with the density functional theory-based molecular dynamics and Kubo-Greenwood approach to unravel the thermal evolution of Earth's core. Furthermore, the Ni effect on the thermal evolution of Earth's core was systematically calculated and analyzed, including the age of Earth's inner core and detailed thermal stratification of Earth's outer core.

## 2. Methods
### 2.1. Thermal evolution of the Earth's core

Previous studies have conventionally assumed that vigorous convection produces an outer core that is in a well-mixed, adiabatic state on the timescales of interest. By employing the Boussinesq approximation, the general energy budget for Earth's core (Buffett, Huppert et al. 1996, Labrosse, Poirier et al. 2001, Gubbins, Alfe et al. 2003, Roberts, Jones et al. 2003, Gubbins, Alfe et al. 2004, Davies and Gubbins 2011, Nimmo 2015) is as summarized. Thermal (Gubbins, Alfe et al. 2003) and compositional convection (Gubbins, Alfe et al. 2004) were provided in the energy and entropy budget of the Earth's core, aside from the possibility of stratified layers. The total amount of energy extracted from the core (core-mantle boundary heat flow $Q_{CMB}$) depends on the radioactive heat production ($Q_R$), secular cooling ($Q_S$), contraction of the core ($Q_P$), heat of reaction ($Q_H$), chemical gravitational ($Q_g$), and latent heat ($Q_L$) release with the growth of the inner core, expressed as follows:

$$Q_{CMB} = Q_R + Q_s + Q_P + Q_H + Q_g + Q_L, \tag{1}$$

where $Q_P$ and $Q_H$ are neglected. This equation demonstrates that neither the adiabatic heat flow ($Q_{ad}$) nor Ohmic heat play any role in the global energy budget. The $Q_{ad}$ parameter can be expressed as follows:

$$Q_{ad} = -4\pi r^2 k \left(\frac{\partial T}{\partial r}\right)_s, \tag{2}$$

where k is the thermal conductivity and $\frac{\partial T}{\partial r}$ is the temperature gradient. Thermal convection occurs when the total heat flow is higher than the adiabatic heat flow, i.e. $Q_{CMB} > Q_{ad}$.

Neglecting $E_\alpha$ and $E_P$, the entropy budget of Earth's core can be expressed as follows:

$$E_s + E_R + E_L + E_g = E_k + E_\Phi, \tag{3}$$

where E represents entropy and the subscripts s, R, L, and g are the different contributions to the entropy, which have the same meaning as expressed in Eq. (1). Here, $E_k$ depends on the adiabat at the CMB, expressed as follows:

$$E_k = \int k \left(\frac{\nabla T_a}{T_a}\right)^2 dV. \tag{4}$$

The entropy production rate, $E_\Phi$, available to drive the dymamo can be written as follows:

$$E_\Phi = E_R + E_s + E_L + E_g - E_k = E_R + \tilde{E}_T \frac{dT_c}{dt} - E_k. \tag{5}$$

By combining Eqs. (4) and (5), the core heat flow required to sustain a dynamo is characterized by a particular entropy production rate:

$$Q_{CMB} = Q_R \left(1 - \frac{\tilde{Q}_T}{\tilde{E}_T} \frac{1}{T_R}\right) + \frac{\tilde{Q}_T}{\tilde{E}_T}\left(E_\Phi + E_k\right), \tag{6}$$

where $T_R$ is the effective temperature, such that $T_R = Q_R / E_R$.

The age of the inner core has a close correlation with the thermal conductivity and heat flow at the core-mantle boundary. The age can be calculated based on the limitations of the parameters of the geodynamo. In this paper, we mainly illustrate the two principal limitations, i.e., the limitations on the entropy production rate available to drive the geodynamo (Nimmo 2015) and the net dissipation rate $Q_D$ (Roberts, Jones et al. 2003). Based on the entropy production rate limitation (Nimmo 2015), the age of the inner core $\tau$ is as follows:

$$\tau = \frac{W_s + W_g + W_L + W_R}{Q_{CMB}}, \qquad (7)$$

where the total amount energy released derives from four main sources: secular cooling of the entire core ($W_s = 1.6 \times 10^{28}$ J), gravitational energy ($W_g = 4.4 \times 10^{28}$ J), latent heat release ($W_L = 6.9 \times 10^{28}$ J), and radioactive decay ($W_R$ ignored). As geomagnetism existed prior to the inner core, we considered it to exist all the time. The entropy production rate available to drive the dynamo should be positive. The rate of the entropy production rate within the dynamo ultimately depends on the core-mantle boundary heat flow. The available core-mantle boundary heat flow is 5 TW < $Q_{CMB}$ < 15 TW (Nimmo 2015).

In contrast, Joule losses are related to the creation of the observed geomagnetic field, dominating the dissipation processes, where the range of $Q_D$ is 1 TW < $Q_D$ < 2 TW (Roberts, Jones et al. 2003). By supposing $\dot{M}_{ic} = M_{ic} / (\lambda \times 1.2 Gyr)$ ($M_{ic}$ is the mass of the inner core), $Q_D$ can be expressed as follows:

$$Q_D = \frac{T_D}{T_{CMB}}[(H_{ICB} + Q_L)(1 - \frac{T_{CMB}}{T_{ICB}}) + (Q_s + Q_R)(1 - \frac{T_{CMB}}{T_{ICB}}) + Q_g - \sum T_{CMB}], \qquad (8)$$

where T is temperature (K) at each condition, the subscript CMB and ICB represent the core-mantle and inner core boundaries, respectively, and $T_D$ is the average temperature of inner and outer core. Based on the limitations of Joule losses, the age of the inner core is $\tau = \lambda \times 1.2 Ga$.

## 2.2. Thermal stratification

When $Q_{ad} > Q_{CMB}$, there is a subadiabatic condition in the real outer core, with the occurrence of a stably stratified region. Thermal stratification was analyzed using the effective temperature gradients (Davies and Gubbins 2011), without the consideration of compositional stratification (Buffett and Seagle 2010). A negative value for the effective temperature gradient, dT/dr (Davies and Gubbins 2011), indicates a stably stratified region. In the absence of the inner core, the total temperature gradient is $T' = T'_L + T'_s + T'_g + T'_r - T'_{ad}$. The subscripts L, s, g, r, and ad represent the different contributions to the temperature gradient (Davies and Gubbins 2011):

$$T'_L = \frac{\rho_i L}{\tau_r k_T} \frac{T_i}{T_o} \frac{dT_o}{dt} \frac{r_i^2}{r^2},$$

$$T'_s = -\frac{1}{\kappa_T (r_o^2 - r_i^2)} [\frac{1}{3}(T_i r_o^2 - T_o r_i^2)r - \frac{1}{5}(T_i - T_o)r^3] \frac{1}{T_o} \frac{dT_o}{dt},$$

$$T'_g = -\frac{1}{3\kappa_T} \frac{\alpha_c}{\alpha_T} \frac{4\pi r_i^2 \rho_i c_0}{\tau_r M_{oc}} \frac{T_i}{T_o} \frac{dT_o}{dt} (r - \frac{r_o^3}{r^2}), \qquad (9)$$

$$T'_r = \frac{q_r}{3k_T} r,$$

$$T'_{ad} = -2\frac{k}{k_T} \frac{T_i - T_o}{r_o^2 - r_i^2} r,$$

where $\frac{dT_o}{dt}$ is the cooling rate at the core-mantle boundary (K/Ga), $\rho$ is the density (g/cm$^3$), r is the radius (km), sub-indices o and i are the outer and inner core, respectively, $\kappa_T$ is the thermal diffusivity, $k_T$ is the thermal conductivity, $k_T = \bar{\rho} C_P \kappa_T$, $\alpha_c$ is the compositional expansion coefficient (1.10), $\alpha_T$ is the coefficient of thermal expansion ($1.35 \times 10^{-5} K^{-1}$), $\tau_r$ is the difference between the melting and adiabatic gradients ($-1.66 \times 10^{-4} Km^{-1}$), and $q_r$ is volumetric heat source (Wm$^{-3}$).

## 2.3. Calculation details

Quantum molecular dynamic calculations were performed using the Vienna *ab initio* simulation package (VASP) (Kresse and Hafner 1993) with the constant *NPT* ensemble. The ion-electron interaction was represented by the projector augmented wave (Blöchl 1994) pseudopotential. The Fe_pv and Ni_pv pseudopotential were considered for Fe and Ni atoms. The generalized gradient approximation (Perdew, Burke et al. 1996) with the Perdew-Burke-Ernzerhof corrections were employed. The electronic states were populated following the Fermi-Dirac distribution (Mermin 1965) while the thermodynamic equilibrium of the ions was sustained by setting the value of the electronic temperature $T_e$ identical to that of the ionic temperature $T_i$.

The periodic cubic box containing 128 atoms was used in the calculations. For the FeNi mixtures, 12 atoms of Ni were a randomly adopted in the cell. The time step of the molecular dynamics was set to 1.0 fs. The plane-wave energy cutoff was set to 400 eV. The molecular dynamics simulations were conducted with only Γ-point sampling in the Brillouin zone. A sufficient number of bands were included so that the occupation of the highest band was less than $1 \times 10^{-6}$. The MD simulation was executed for a total of 6,000 steps, where the total energy and pressure fluctuations were less than 1%. The detailed relationship between pressure and temperature along the radial distance at Earth's core condition was obtained from the Preliminary reference Earth model (Dziewonski and Anderson 1981).

Thermal (k) and electrical conductivity (σ) were calculated using the Kubo-Greenwood formula, as implemented in the ABINIT code (Gonze, Beuken et al. 2002, Gonze, Amadon et al. 2009). The electronic current autocorrelation function via Kubo's linear response formalism can be expressed as follows:

$$L_{ij} = (-1)^{i+j} \frac{he^2}{V} \sum_{k',k} \lim_{\varepsilon \to 0} \frac{f(\varepsilon_{k'}) - f(\varepsilon_k)}{\varepsilon} \delta(\varepsilon_{k'} - \varepsilon_k - \varepsilon) \times \langle \psi_k | \mathbf{v} | \psi_{k'} \rangle \langle \psi_{k'} | \mathbf{v} | \psi_k \rangle (\varepsilon_{k'} - \varepsilon_f)^{i-1} (\varepsilon_{k'} - \varepsilon_f)^{j-1},$$

$$\sigma = 1/L_{11},$$

$$k = \frac{1}{e^2 T} \left( L_{22} - \frac{L_{12}^2}{L_{11}} \right).$$

(10)

where $\varepsilon_f$ is the Fermi energy; $\psi_k$, $\varepsilon_k$, and f($\varepsilon_k$) are the wave function, eigenvalue, and Fermi-Dirac occupation of eigenstate k, respectively; $\mathbf{v}$ is the velocity operator; and V is the simulation cell volume.

The thermal and electrical conductivity were obtained by averaging ten snapshots extracted from the last 2,000 time steps in each MD trajectory with an interval of 200 time steps. We note that k and $\sigma$ refer only to the electronic component. As the ionic contribution to thermal conductivity is only a few W/m/K (Pozzo, Davies et al. 2012), the ionic contribution was negligible compared with the electronic contribution.

### 3. Results and Discussion
### 3.1. Thermal and electrical conductivity

Table 1 lists the detailed thermal and electrical conductivities. Our results for pure Fe agree with previously reported ab initio simulation results, where the thermal conductivity of Fe along the adiabatic curve ranges from 125.07 to 216.18 W/m/K, as shown in Figure 1(a). As illustrated previously (de Koker, Steinle-Neumann et al. 2012, Pozzo, Davies et al. 2012), our calculated thermal conductivity is more or less higher than the experimental results. However, the addition of Ni reduces both the thermal and electrical conductivity relative to their values of pure Fe at the same pressure and temperature at the condition throughout the entirety of Earth's core. Nickel reduces the thermal conductivity to, on average, ~12.30 W/m/K at the conditions present throughout the entirety of Earth's core. Nickel reduces the electronic conductivity, on average, to ~957 $(\Omega cm)^{-1}$, which is consistent with previously reported resistivity experimental results on Fe-Ni alloys at the core-mantle and inner-core boundary conditions (Gomi and Hirose 2015) at several GPa (Pommier 2020).

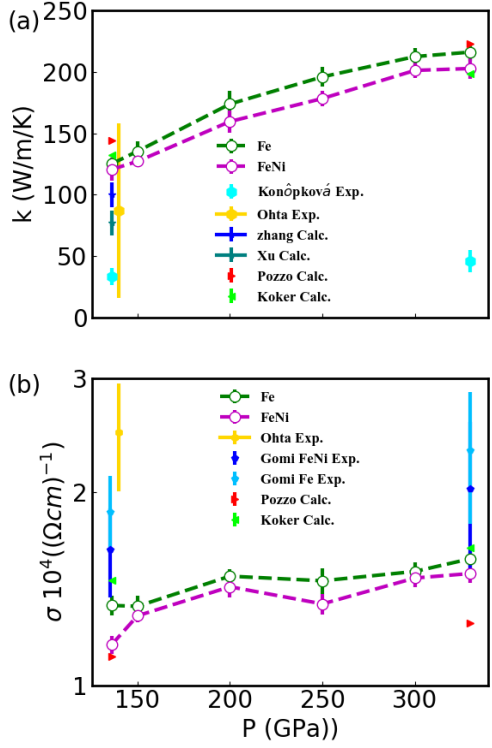

Figure 1 (a) Thermal (k, W/m/K) and (b) electrical ($\sigma$ $10^4 (\Omega cm)^{-1}$) conductivity as a function of pressure (P, GPa) along the adiabatic curves at Earth's core conditions. The green and purple line represent the results for Fe and FeNi mixture, respectively. The scatter data were obtained from previous studies (de Koker, Steinle-Neumann et al. 2012, Pozzo, Davies et al. 2012, Gomi and Hirose 2015, Konôpková, McWilliams et al. 2016, Ohta, Kuwayama et al. 2016, Xu, Zhang et al. 2018, Zhang, Hou et al. 2020). 'Exp.' and 'Calc.' represent the experimental and theoretical results, respectively. 'Konôpková Exp.' from (Konôpková, McWilliams et al. 2016): direct measurements of thermal conductivity via a laser-heated diamond-anvil cell. 'Ohta Exp.' from (Ohta, Kuwayama et al. 2016) for direct resistivity measurements via a laser-heated diamond-anvil cell . 'Gomi Fe Exp.' and 'Gomi FeNi Exp'. both from (Gomi and Hirose 2015) for experiments on hcp Fe and Fe-Ni alloys, respectively. 'Zhang Calc.' from (Zhang, Hou et al. 2020) via a shocked experiment. 'Xu Calc.' from (Xu, Zhang et al. 2018), which considered the electron-electron scattering effect. 'Pozzo Calc.' from (Pozzo, Davies et al. 2012) and 'Koker Calc.' from (de Koker, Steinle-Neumann et al. 2012), both of which performed ab initio simulations.

We emphasize that the thermal (k) and electrical ($\sigma$) conductivities can be directly evaluated without using any assumptions with respect to the Lorenz number. The Lorenz number is highly dependent on two nondimensional parameters, i.e., the degeneracy, $\theta$, and coupling parameters, $\Gamma$. Based on the law of Wiedemann and Franz (Chester and Thellung 1961), the Lorenz number can be defined as follows:

$$L = \frac{k}{\sigma T} = \gamma \frac{k_B^2}{e^2}, \tag{11}$$

where $\gamma$ is a dimensionless parameter that depends on the screened potential and corresponds to the scattering of electrons (Chester and Thellung 1961). In a degenerate ($\theta \leq 1$) and coupled plasma ($\Gamma \geq 1$), $\gamma$ reaches the ideal Sommerfeld number (constant $\pi^2/3$), which is the value valid for metals. In this situation, the Wiedemann-Franz law recovers in an elastic interaction electron system. In a nondegenerate case ($\theta \geq 1$ and $\Gamma \leq 1$), as the temperature is high, $\gamma$ reaches the value of kinetic matter (4 or 1.5966 depending on the electron-electron collisions). In the intermediate region, there are no assumptions for predicting the Lorenz number and the thermal conductivity cannot be deduced from the electrical conductivity using the Wiedemann-Franz law. At Earth's core condition, the Lorenz number $\gamma$ is notably near the Sommerfeld limit and its departure from the ideal value.

In this case, Ni has a nonnegligible effect on the thermal evolution of Earth's core. As the thermal and electrical conductivities of pure Ni are higher than pure Fe at identical conditions in Earth's core, the simple mixing rules (or Matthiessen's rule) are no longer valid. As Ni lies in the vicinity of Fe in the Periodic Table, its effect on conductivity is not as large as lighter elements, such as S and Si. We can illustrate the effects of Ni as follows. The atomic mass and valence electrons of Ni are larger than Fe. The volume of the Ni system is larger than the corresponding Fe system at the same pressure and temperature. The effective electronic density (or electron ionization) of FeNi mixtures is larger than pure Fe. Owing to the total effect of the larger ionic and electronic properties, the electron ionization per volume of a FeNi mixture is lower than pure Fe, which results in lower thermal and electrical conductivities. This difference decreases with a decrease in the pressure and temperature. In contrast, the lower thermal and electrical conductivities of FeNi systems may originate from the Ni doping effect on the electron structure, the randomness or clustering of Ni elements, and so on. A more detailed analysis on the origin of the Ni effect is not in the scope of this paper.

Table 1 The calculated thermal and electrical conductivities for pure Fe and FeNi mixture along the adiabatic curve at Earth's core condition. The Ni concentration in the FeNi mixture is 9.375%

| P GPa | K W/m/K | | $\sigma$ $(\Omega\ cm)^{-1}$ | |
|---|---|---|---|---|
| | Fe | FeNi | Fe | FeNi |
| 136 | 125.07 (5.84) | 120.52 (9.47) | 13308 (654) | 11560 (360) |
| 150 | 135.31 (7.82) | 127.06 (4.26) | 13270 (523) | 12824 (286) |
| 200 | 173.88 (10.66) | 159.59 (9.20) | 14780 (393) | 14229 (543) |
| 250 | 196.02 (8.11) | 178.36 (6.46) | 14539 (668) | 13382 (502) |
| 300 | 212.52 (6.70) | 201.28 (6.38) | 15032 (504) | 14683 (485) |
| 330 | 216.18 (6.95) | 202.80 (8.62) | 15733 (402) | 14921 (499) |

### 3.2. Age of the inner core

The heat flow at the core-mantle boundary is crucially important for both the compositional and thermal convection mechanisms and significantly influences the geodynamo. Furthermore, the heat flow at the core-mantle boundary is closely related to the recreation of the inner core. Currently, the accepted core-mantle boundary heat flow is 12 ± 5 TW (Nimmo 2015). The age of the inner core was calculated from the energy and entropy budget of Earth's core, with restrictions from the rate of entropy production for the dynamo (Nimmo 2015) or Joule losses that dominate the dissipative process (Roberts, Jones et al. 2003). Our results agree are consistent with recently reported high thermal conductivities, which suggest a younger inner core age within 1 Ga (Pozzo, Davies et al. 2012, Gomi, Ohta et al. 2013, Nimmo 2015).

In Figure 2(a), the lines of the core-mantle boundary heat flow as a function of the entropy production rate are almost linear and the difference between pure Fe and FeNi mixtures is 0.411 TW. When the entropy production rate is ~600 MW/K, the corresponding core-mantle boundary heat flow is ~12 TW, with an age of the inner core for FeNi mixtures and pure Fe of ~0.605 Ga. The effect of Ni on the age of the inner core is only 0.0004 Ga longer, within an increase of 0.07%. When the heat flow at the core-mantle boundary increases to 15 TW, the age of FeNi mixtures corresponds to 0.484 Ga. When the core-mantle boundary heat flow is lower, the difference between the pure Fe and FeNi mixtures may be large. The upper limit for the age is 0.67 Ga for pure Fe systems and 0.69 Ga for FeNi mixtures, which corresponds to a zero entropy production rate while the Ni effect induces a 2.99% increase in the age of the inner core. If we suppose that the entropy production rate is only 100 MW/K, the age difference between pure Fe and FeNi mixtures is 0.072 Ga. With the limitation of Joule losses within (1TW, 2TW) in Figure 2(b), the age of the inner core ranges from 0.653 to 0.884 Ga for pure Fe and 0.663 to 0.929 Ga for FeNi mixtures. The FeNi mixture is 0.045 Ga longer (5.13% increase) than Fe with 1 TW Joule losses. The FeNi mixture is 0.018Ga longer (3.20% increase) than Fe with 2 TW Joule losses. The heat flow at the core-mantle boundary for FeNi mixtures is 0.103 TW lower than pure Fe with the same Joule losses.

These two methods coincide with each other, but the specified age values are different from each other within the scale of limitations. The proposed Joule losses were previously underestimated, such that the age of the inner core is large compared with recently reported results. The larger heat flow at the core-mantle boundary, the small difference in the ages between the FeNi mixtures and pure Fe. In other words, the effect that Ni has on the age of the inner core should be considered when heat flow at the core-mantle boundary is lower. The older inner core corresponds to a slow cooling core, i.e., the slightly lower thermal conductivity. High thermal conductivity corresponds to a big age of the inner core. The addition of Ni with a decrease of ~10 W/m/k leads to an age for the inner core of at most 0.2 Ga younger than a pure Fe inner core.

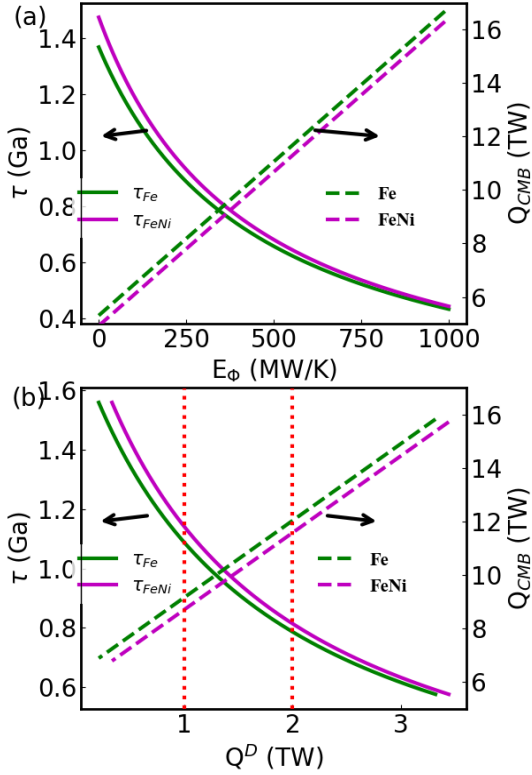

Figure 2 Calculations for the age of the inner core ($\tau$, Ga) as a function of both the (a) entropy production rate ($E_\Phi$, MW/K) and (b) Joule loss ($Q_D$, TW) limitations. In addition, the heat flow at the core-mantle boundary ($Q_{CMB}$) is also calculated. The entropy for the dynamo should be positive ($E_D > 0$) (Nimmo 2015) in (a). Joule losses are in the range of 1 TW < $Q_D$ < 2 TW (Roberts, Jones et al. 2003), which are denoted by the red dotted lines in (b). Green lines correspond to pure Fe systems while purple corresponds to FeNi systems. The lines correspond to the age of the inner core as a function of the (a) entropy production rate and (b) Joule losses. The long dotted lines correspond to the heat flow at the core-mantle boundary as a function of the (a) entropy production rate and (b) Joule losses.

### 3.3. Stable thermal stratification

When the heat flow at the core-mantle boundary is less than the adiabatic heat flow at the core-mantle boundary, thermal convection does not exist throughout Earth's outer core; rather, there is the presence of stratification. Stable stratification may alter the characteristics of the geodynamo. The adiabatic heat flow of pure Fe ranges from 1.68 to 15.64 TW while the FeNi mixture ranges from 1.58 to 15.18 TW, as shown in Figure 3(a). We note that the heat flow at the core-mantle boundary is less than the adiabatic heat flow for both pure Fe and FeNi. We analyzed the temperature gradient, where a negative value indicates a stably-stratified region (Davies and Gubbins 2011).

Although the adiabatic heat flow monotonically increases with distance, the adiabatic temperature gradient has a maximum of ~2,800 km for Fe and 2,828 km for FeNi mixtures (Figure 3(a)). Temperature gradients $T_s'$, $T_g'$, and $T_L'$ correlate with the cooling rate at the core-mantle boundary (dT$_o$/dt). Stratification exists on top of the outer core, as shown in Figure 3(b). When dT$_o$/dt = 126 K/Ga, the thickness of the stable layer is 417.02 km for pure Fe and 320.12 km for FeNi mixtures. The thermal stratification retreated with the appearance of Ni in the outer core, as shown in Figure 3(c). When dT$_o$/dt > 170 K/Ga, thermal stratification disappears. Here, $dT_o/dt$ has a direct correlation with the heat flow at the core-mantle boundary. The high heat flow at the core-mantle boundary corresponds to a rapid decrease in the temperature at the core-mantle boundary. Thermal convection does not occur throughout the outer core. This indicates that the energy can be conducted along the adiabatic curve, such that eddy or turbulent motion is not required inside the outer core, resulting in the presence of a stratification layer.

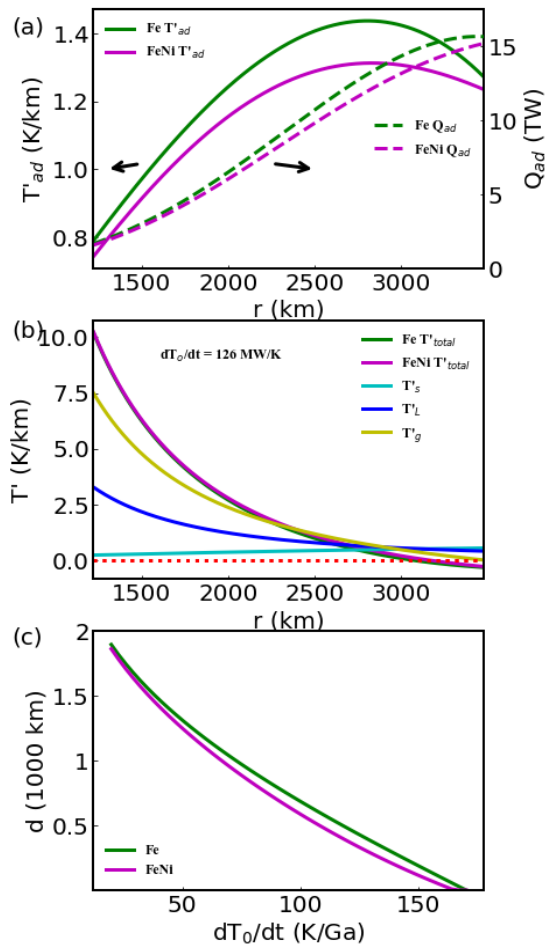

Figure 3 Thickness of the thermal stable stratification in Fe and FeNi mixtures by analyzing the effective temperature gradient. (a) Adiabatic temperature gradients (T'$_{ad}$, K/km) and heat flow as a function of the radial distance (r, km). (b) Detailed correlations between temperature gradients and distance when dT$_o$/dt = 126 MW/K. T'$_{total}$, T'$_s$, T'$_L$, and T'$_g$ represent total effective, secular cooling, latent heat, and compositional buoyancy temperature gradient, respectively. The red dotted line corresponds to T' = 0 K/km. (c) The thickness of the stable stratification as a function of the rate of temperature drop at the core-mantle boundary (dT$_o$/dt, K/Ga). Green lines correspond to pure Fe while purple corresponds to FeNi mixtures in (a) and (c). The long dotted lines in (a) correspond to the relationship between the adiabatic heat flow and radial distance.

## 4. Conclusions

The calculations presented here are the first obtained for the thermal conductivity of FeNi mixtures at Earth's outer core condition through ab initio simulations. Nickel is the second main element in the core and is of significant importance for the thermal evolution of Earth's core. We systematically analyzed the effect of Ni on the thermal conductivity, age of the inner core, and thickness of the thermal stratification in the outer core based on a combination of ab initio simulations and the thermal evolution theory of Earth's core.

The ab initio simulation results show that Ni reduces the thermal (12.30 W/m/K on average) and electrical conductivities (957 $(\Omega cm)^{-1}$ on average) along the adiabatic curve throughout the entirety of Earth's outer core. Based on the energy and entropy budget of Earth's core, it is plausible that the age of the inner core is ~0.60-0.70 Ga as a result of the thermal conductivity of the FeNi mixture. The existence of Ni in Earth's core can increase the age of the inner core by a maximum of 0.20 Ga. By analyzing the temperature gradients, the addition of Ni results in a reduction in the thickness of the thermal stratification in Earth's outer core from 417.02 to 320.12 km when $dT_o/dt$ = 126 K/Ga. Based on observations of heat flow at the core-mantle boundary, the effect that Ni has on the age of the inner core decreases while the thickness of thermal stratification increases with a high heat flow at the core-mantle boundary. Compared with only considering pure Fe, the selected FeNi mixtures are more closely related to the actual conditions in Earth's core. Taking the effect of Ni into consideration for the thermal evolution of Earth's core enables us to obtain a more credible assessment of Earth's core, which is also of great significance for geomagnetic and geodynamo modeling.

**Acknowledgments**

We wish to acknowledge support from the National key R&D program of China (Grant No. 2017YFA0403200), the National Natural Science Foundation of China (NSFC) (Grant Nos. 11975058, 11775031 and 11625415), the Foundation for the Development of Science and Technology of the China Academy of Engineering Physics, and Science Challenge Project under Grant No. TZ2016001.